\documentclass[journal,twoside,web]{ieeecolor}
\usepackage{jsen}
\usepackage{cite}
\usepackage{amsmath,amssymb,amsfonts}
\usepackage{algorithmic}
\usepackage{graphicx}
\usepackage{textcomp}
\usepackage{wrapfig}
\usepackage{hyperref}
\def\BibTeX{{\rm B\kern-.05em{\sc i\kern-.025em b}\kern-.08em
    T\kern-.1667em\lower.7ex\hbox{E}\kern-.125emX}}
\definecolor{abstractbg}{rgb}{0.89804,0.94510,0.83137}
\begin{document}
\title{Multi-Frequency Impedance Myography:\\The PhaseX Effect}
\author{Roman~Kusche,
        and~Martin~Ryschka
\thanks{Manuscript received Month ??, 2020; revised Month ??, 2020; accepted Month ??, 2020. Date of publication Month ??, 2020; date of current version Month ??, 2020. This work was supported by the German Federal Ministry of Education and Research (BMBF) under the project INOPRO (FKZ16SV7666). \textit{(Corresponding author:
	Roman Kusche.)}}
\thanks{R. Kusche is with the Center of Excellence CoSA, Luebeck University of Applied Sciences, Luebeck 23562, Germany (e-mail:roman.kusche@th-luebeck.de).}
\thanks{M. Ryschka is with the Luebeck University of Applied Sciences.}
}

\IEEEtitleabstractindextext{%
\fcolorbox{abstractbg}{abstractbg}{%
\begin{minipage}{\textwidth}%
\begin{wrapfigure}[12]{r}{3in}%
\includegraphics[width=2.9in]{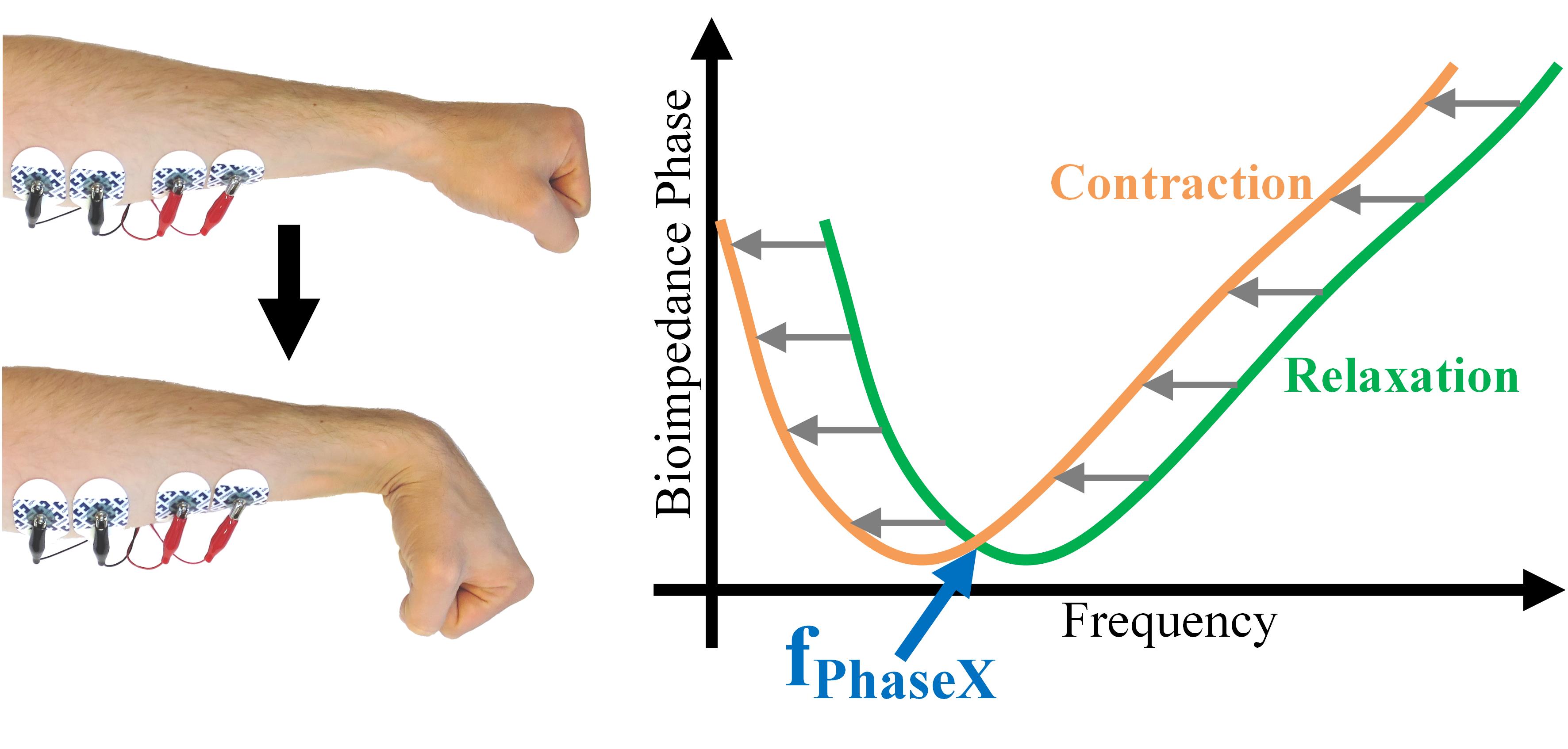}%
\end{wrapfigure}%
\begin{abstract}
Muscle contraction is often detected via EMG in prosthetics. However, signal disturbances due to electrode motions can lead to misinterpretations. Therefore, alternative measurement approaches are desired to increase the reliability of the results.
In this work, a novel approach based on impedance myography is proposed. By means of an equivalent circuit of a muscle, its electrical characteristics during contractions are analyzed. In this analysis, a new biomedical marker named the PhaseX Effect is described. This effect is based on the specific behavior of the phase response when the muscle is contracted and is interesting due to its high signal robustness and low signal processing requirements. The resilience of this effect against electrode motion is also analyzed.
Measurements of the complex impedance myography spectra are performed on the forearms of three subjects during relaxation and contraction of the corresponding muscle.
The subject measurements show the expected behavior of the muscle model. Actual muscle contractions can easily be detected via a simple analysis of the phase response. For a better visualization, the measurements are repeated while acquiring a synchronized video of the moving forearm.
The particular effect of the phase response during muscle contraction can be used as a new marker that can be beneficial in several applications such as prostheses control. The PhaseX Effect has high reliability and low signal processing requirements, making it advantageous over other muscle activity markers.
The combination of a reliable marker and simple signal analysis promises to become a new method for prostheses control.
\end{abstract}

\end{minipage}}}

\maketitle

\section{Introduction}

\IEEEPARstart{M}{uscle} contraction detection is widely used in several human-machine interaction applications and is not exclusively of interest for biomedical purposes \cite{Wheeler2006,Bitzera,Kiguchi2004,Sun2020}.
However, many approaches strive to achieve active control of limb prostheses.
Due to the possibility of measuring muscle action potentials directly via surface electromyography (EMG), this technique has become very popular \cite{Esposito2018}.
Furthermore, the electrical characteristics of EMG signals are technically comfortable, with frequencies in the hundreds of Hertz and amplitudes in the $\mathrm{mV}$ range, which simplifies the development of the required EMG-specific electronic circuitry \cite{Moxham1982,WileyAssistiveTechnology}.
To connect these electronic circuits to the human body, electrodes are utilized to interfacing the corresponding conductions by electrons and ions.
The behavior of the resulting electrochemical half-cells at the voltage-measurement electrodes has been extensively investigated in previous studies \cite{BioelectricityandBioimpedanceBasics}.
In EMG measurements, especially their specific half-cell voltages, which are electrically in series to the EMG source, can be challenging.
Even if these voltages are DC voltages, mechanical disturbances, such as vibrations at the electrode sites, can cause temporary changes, leading to AC signal components.
These disturbances are superimposed on the EMG signals and can produce signal components of similar frequency and amplitude ranges as the actual EMG \cite{Roland2017}. Due to this, signal misinterpretation is possible.
Since, especially in biomedical applications, these misinterpretations can provoke dangerous situations such as undesired prosthesis movements, alternative or redundant approaches would be beneficial.

\begin{figure}[!t]
	\centering
	\includegraphics[width=0.48\textwidth]{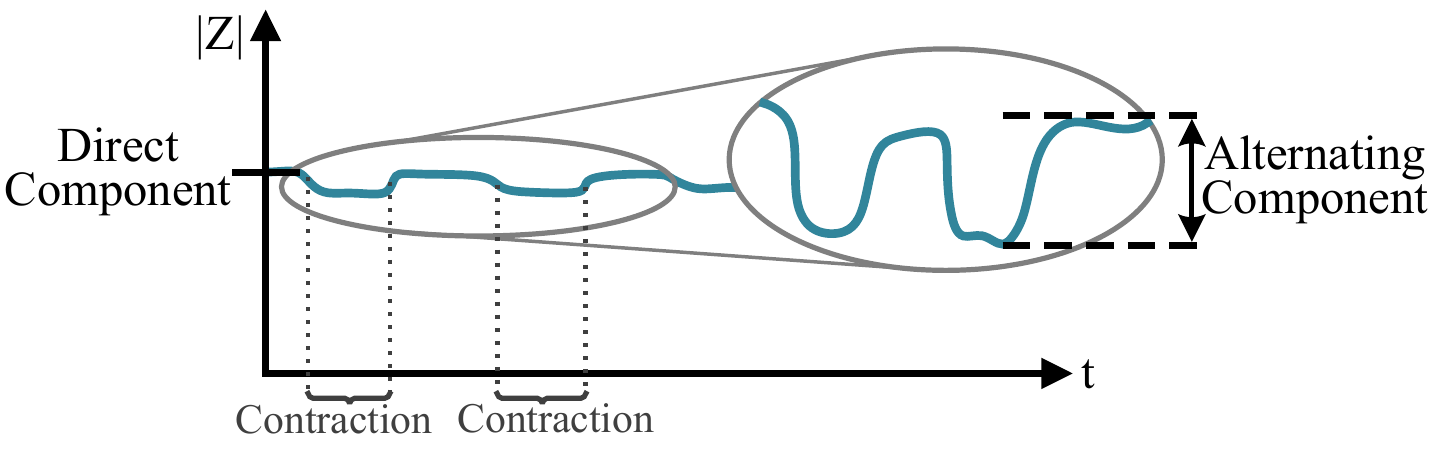}
\caption{Illustration of the transient bioimpedance magnitude over time during muscle contractions. The characteristic signal morphology is typically not significantly affected by the choice of excitation frequency. Based on published measurements \cite{Rutkove2009,CombiningBioimpedanceandEMGMeasurementsforReliableMuscleContractionDetection}.}
	\label{ImpedanceMyographieSignalExample}
\end{figure}

In addition to mechanical or optical measurement setups, the passive electrical characteristics of muscles provide information regarding contractions and relaxations \cite{Esposito2018,Bifulco2017,Zhang2014,Chianura2010,Bianchi1999}.
The determination of the electrical bioimpedance of muscles or muscle regions is called electrical impedance myography (EIM) and is still a niche field in biomedical engineering \cite{Shiffman2003,Rutkove2009,Hornero2013}.
In Fig. \ref{ImpedanceMyographieSignalExample}, an example impedance myography signal is illustrated in the time domain. For this simplified explanation, only the impedance magnitude is shown.
It consists of a direct component, corresponding to the mean bioimpedance magnitude of the investigated tissue under examination, and an alternating component that is caused by the muscle contraction and its resulting geometrical changes \cite{Rutkove2009,CombiningBioimpedanceandEMGMeasurementsforReliableMuscleContractionDetection}.

To determine the bioimpedance, a small known alternating excitation current is applied to the tissue. Measuring the resulting voltage drop enables the calculation of the complex impedance \cite{U2004}.
The major advantage of this active technique is the possibility of applying excitation currents with known frequencies that are much higher than those of typical disturbances \cite{BioelectricityandBioimpedanceBasics}.
Demodulation of the measured voltage signal with knowledge of the excitation frequency allows separation of the desired signal from noise and disturbances \cite{Min2000}.
However, the complexity of applicable measurement setups is much higher than those of common EMG circuits.
Furthermore, both actual muscle contractions and mechanical deformations and blood perfusion of the tissue region cause changes in electrical bioimpedance \cite{Rutkove2009}.
It is therefore desirable to find signal markers that are exclusively sensitive to muscle contractions.

In this work, we present a novel simple biosignal marker that reliably indicates actual muscle contractions via multi-frequency electrical impedance myography and clearly separates them from external disturbances. For better readability, we name the specific behavior the PhaseX Effect.
Following a brief introduction to complex electrical impedance myography, we explain the theoretical basics of our new approach using an equivalent circuit for skeletal muscles.
Afterward, we discuss the influence of motion artifacts on real electrical impedance myography measurements and how they differ from the behavior of actual muscle contractions.
Finally, the theoretical approach is supported by subject measurements.


\section{Materials and Methods}

\subsection{Impedance Myographic PhaseX Effect} \label{sec:EIMphaseX}
In contrast to electromyography, electrical impedance myography is not based on the measurement of muscle action potentials.
Instead, bioimpedance measurements are utilized to detect the corresponding geometrical changes of the muscles and surrounding tissue \cite{Rutkove2009}.
As mentioned before, the major advantage of electrical impedance myography over EMG is that the frequency of the bioimpedance measurement signal is significantly higher than that of motion artifacts.
In addition, this frequency is known and therefore enables a simple distinction between wanted and interfering signals.
Furthermore, the possibility of choosing and changing the excitation frequency allows us to determine the bioimpedance for several frequencies and thus to determine its complex frequency response.

\begin{figure}[!t]
	\centering
	\includegraphics[width=0.48\textwidth]{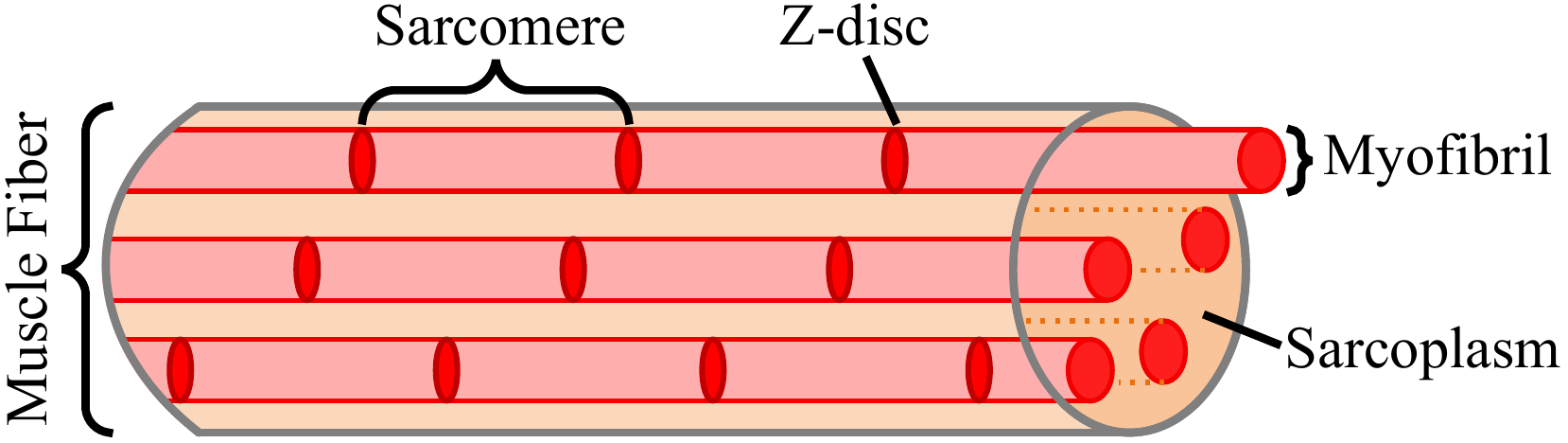}
\caption{Simplified geometrical model of a skeletal muscle fiber, based on \cite{PrinciplesofAnatomyandPhysiology}. }
	\label{MuscleFibre}
\end{figure}

Skeletal muscles consist of various muscle fiber bundles, which are each composed of many individual muscle fibers.
Each fiber consists of myofibrils surrounded by sarcoplasm, as illustrated in Fig. \ref{MuscleFibre} \cite{PrinciplesofAnatomyandPhysiology}.
These myofibrils in turn consist of sequences of many sarcomeres, which are separated by so-called Z-discs.
Based on this model, it is assumed that the sarcoplasm, which mostly consists of water and proteins, has resistive electrical behavior \cite{PrinciplesofAnatomyandPhysiology}.
The sarcomeres in parallel are also considered to be resistive.
The alternating structure of sarcomeres and Z-discs is assumed to cause capacitive behavior within the skeletal muscle fiber \cite{PrinciplesofAnatomyandPhysiology}.
Due to geometrical anisotropy, these simplifications are only valid for bioimpedance in the muscle fiber direction \cite{Matthie2008}.

\begin{figure}[!t]
	\centering
	\includegraphics[width=0.3\textwidth]{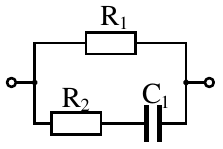}
\caption{Equivalent circuit of the model depicted in Fig. \ref{MuscleFibre}. It is assumed that $R_\mathrm{1}$ represents the resistive behavior of the sarcoplasm. $R_\mathrm{2}$ and $C_\mathrm{1}$ in series represent the alternating sarcomere and Z-discs. Due to the anisometric geometry, this approach is only applicable to the bioelectrical behavior in the fiber direction.}
	\label{MuscleFibreESB}
\end{figure}

Fig. \ref{MuscleFibreESB} shows the resulting equivalent circuit diagram, in which $R_\mathrm{1}$ represents the sarcoplasm, $R_\mathrm{2}$ the resistive and $C_\mathrm{1}$ the capacitive behavior of the myofibrils.

The bioimpedance of the modeled muscle $Z_\mathrm{Muscle}$ can thus be determined via

\begin{equation}
    Z_\mathrm{Muscle}=\frac{R_\mathrm{1}+\mathrm{j}\omega R_\mathrm{1}R_\mathrm{2}C_\mathrm{1}}{1+\mathrm{j}\omega C_\mathrm{1}(R_\mathrm{1}+R_\mathrm{2})},
\end{equation}

where $\text{j}$ is the imaginary unit and $\omega$ represents the frequency. This results in an impedance magnitude of

\begin{align}
\text{\begin{small}
$|Z_\mathrm{Muscle}| \negthickspace=\negthickspace \frac{ \sqrt{ \left( R_\mathrm{1} \negthickspace + \negthickspace \omega^2 R_\mathrm{1} R_\mathrm{2} C_\mathrm{1}^2 (R_\mathrm{1}+R_\mathrm{2})\right) ^2  \negthickspace +\negthickspace \left( \omega R_\mathrm{1}^2 C_\mathrm{1} \right)^2    } } {    1 \negthickspace+\negthickspace \omega^2 C_\mathrm{1}^2(R_\mathrm{1} \negthickspace+\negthickspace R_\mathrm{2})^2    } .$
\end{small}}
\label{equ:BioZmuskelBetrag}
\end{align}

Contractions of the muscle lead to shortening of the sarcomeres and thus to reductions in the distances between the Z-discs \cite{PrinciplesofAnatomyandPhysiology}.
This increases the capacitance $C_\mathrm{1}$, which corresponds to a decrease in the electrical reactance. Since the combination of $R_\mathrm{1}$, $R_\mathrm{2}$ and $C_\mathrm{1}$ in Fig. \ref{MuscleFibreESB} results in a passive first-order circuit, the impedance magnitude drops over the whole frequency range.

Given the bioimpedance phase

\begin{align}
\text{\begin{small}
$\phi(Z_\mathrm{Muscle})=\mathrm{-arctan}\left(  \frac{\omega C_\mathrm{1} R_\mathrm{1}^2 }{R_\mathrm{1}+\omega^2 C_\mathrm{1}^2 R_\mathrm{1} R_\mathrm{2} \left(R_\mathrm{1}+R_\mathrm{2}\right)}              \right)$,
\end{small}}
\label{equ:BioZmuskelphase}
\end{align}

$\omega$ and $C_\mathrm{1}$ as well as $\omega^2$ and $C_\mathrm{1}^2$ occur only as mathematical products and therefore have the same influence on the phase response.
According to the assumptions given previously that muscle contractions are associated with increasing $C_\mathrm{1}$, this leads to a compression of the phase response in the frequency dimension.
In the logarithmic frequency representation, this corresponds to a shift of the phase response to lower frequencies.
This specific behavior is the theoretical basis of the PhaseX Effect.
The values of $R_\mathrm{1}$ and $R_\mathrm{2}$ are also affected by muscle contraction, but based on previous measurements, it is assumed that these relative changes are smaller than those of $C_\mathrm{1}$.

\begin{figure}[!t]
	\centering
	\includegraphics[width=0.48\textwidth]{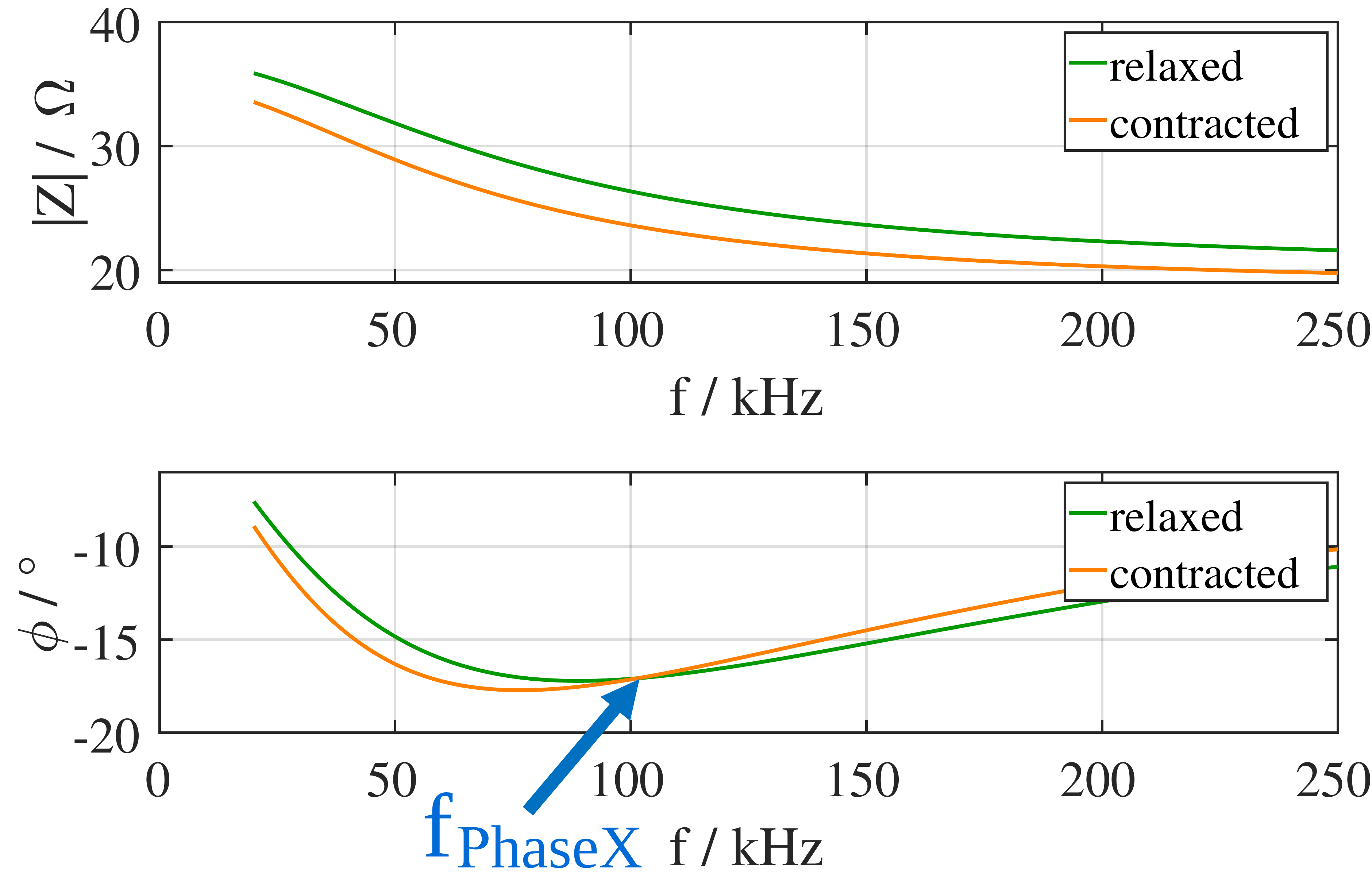}
\caption{Simulated frequency response of the equivalent circuit of a muscle (Fig. \ref{MuscleFibreESB}) in the frequency range of typical bioimpedance measurements. Based on previous measurements, the component values were chosen to be $R_\mathrm{1}=37~\Omega$, $R_\mathrm{2}=44~\Omega$, $C_\mathrm{1}=30~\text{nF}$ during relaxation and $R_\mathrm{1}=35~\Omega$, $R_\mathrm{2}=40~\Omega$, $C_\mathrm{1}=38~\text{nF}$ during contraction.}
	\label{Modell}
\end{figure}

For better visualization of this idea, the equivalent circuit in Fig. \ref{MuscleFibreESB} is simulated using component values that were determined via prior impedance measurements and are comparable to literature values \cite{Nescolarde2013}. These values were $R_\mathrm{1}=37~\Omega$, $R_\mathrm{2}=44~\Omega$ and $C_\mathrm{1}=30~\text{nF}$ during relaxation and $R_\mathrm{1}=35~\Omega$, $R_\mathrm{2}=40~\Omega$, $C_\mathrm{1}=38~\text{nF}$ during contraction.
Both corresponding simulated frequency responses are plotted in Fig. \ref{Modell} in the frequency range typically used for bioimpedance measurements.
In the upper plot, which represents the bioimpedance magnitude response, a decrease over the whole frequency range can be seen when the muscle is contracted.
Below, the phase response is depicted.
It can clearly be seen that utilization of this model leads to a shift of the phase response to lower frequencies during muscle contraction.
Due to this shift, a crossing of the green and orange phase responses occurs at $f_\mathrm{PhaseX} \approx 101~\text{kHz}$.
This crossing is the origin of the principle's name, and for measurement setups, it is assumed to be significantly advantageous.
For reliable recognition of muscle contractions, it thus seems to be sufficient to measure the phase shift with only two different excitation signal frequencies (quasi-)simultaneously.
Choosing these frequencies to be $f_\mathrm{1}<f_\mathrm{PhaseX}$ and $f_\mathrm{2}>f_\mathrm{PhaseX}$ and comparing the transient changes of the corresponding phase values allows simple but meaningful interpretations, as shown in (\ref{eq:phase1}) and (\ref{eq:phase2}).

\begin{align}
\phi(f_\mathrm{1})\downarrow \text{ and } \phi(f_\mathrm{2})\uparrow:  &~~~\text{contraction} \label{eq:phase1}\\
\phi(f_\mathrm{1})\uparrow \text{ and } \phi(f_\mathrm{2})\downarrow:  &~~~\text{relaxation}  \label{eq:phase2}
\end{align}

An alternative evaluation method is the detection of the minimum phase shift and its displacement over frequency.
The actual uniqueness of this behavior will be discussed in the following section.

\subsection{Robustness Against Motion Artifacts}
As mentioned above, the use of known excitation signals in the kHz-frequency range leads to the advantage of simple distinguishing between the actual muscle contractions and motion artifacts. However, this statement is only true for ideal measurement circuits. In this section, we analyze the effect of motion artifacts considering realistic measurement systems and demonstrate the benefit of the proposed PhaseX Effect marker.
The measurement of the bioimpedance is technically challenging.
One particular aspect is that electrode-skin interfaces occur between the actual bioimpedance $Z_\mathrm{Bio}$ and the measuring circuitry \cite{BioelectricityandBioimpedanceBasics}.
It is to be expected that external influences, especially forces and movements acting on the electrodes, can interfere with the bioimpedance measurement.
Since the frequency range of such motion artifacts coincides with that of the muscle contractions, misinterpretations of muscle activity is likely.
However, with a simplified model of the electrode-skin impedances, we will show in this chapter that motion artifacts only cause unidirectional alterations of the bioimpedance phase and can thus be clearly distinguished from actual muscle activity with phase shifts according to (\ref{eq:phase1}) and (\ref{eq:phase2}).

\begin{figure}[!t]
	\centering
	\includegraphics[width=0.48\textwidth]{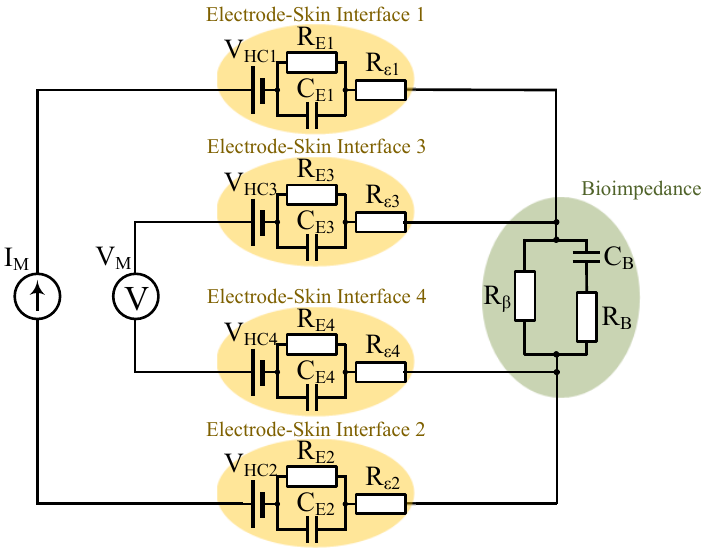}
\caption{Equivalent circuit of an electrical bioimpedance measurement. In this four-terminal setup, electrodes 1 and 2 apply the excitation current $I_\mathrm{M}$ to the investigated tissue, whereas electrodes 3 and 4 are intended for voltage measurement.}
	\label{BioZ_ESB}
\end{figure}

The electrical characteristics of the electrode-skin interfaces can be explained by the combination of the electrochemically caused half-cell voltages ($V_\mathrm{HC}$) and frequency-dependent electrode-skin interface impedances ($Z_\mathrm{E}$) \cite{Dryelectrodesforbioimpedancemeasurementsdesigncharacterizationandcomparison,BioelectricityandBioimpedanceBasics,Marquez2013,Beckmann2010}.
Each of these electrode-skin impedances can be modeled with a combination of the three passive electric components $R_\mathrm{E}$, $C_\mathrm{E}$ and $R_\mathrm{\epsilon}$ \cite{Chi2010, BioelectricityandBioimpedanceBasics}.
Since these complex impedances are in much higher ranges than the actual bioimpedance $Z_\mathrm{Bio}$ of interest, four-terminal measurement setups are typically used.
In Fig. \ref{BioZ_ESB}, the schematic of a typical bioimpedance measurement setup is illustrated.
The excitation current $I_\mathrm{M}$ is applied via electrode-skin interfaces 1 and 2 to the bioimpedance, represented by $R_\mathrm{B}$, $C_\mathrm{B}$ and $R_\mathrm{\beta}$.
This $R||(R+C)$-circuit is similar to the equivalent circuit of a muscle fiber in Fig. \ref{MuscleFibreESB}, but $Z_\mathrm{Bio}$ comprises the whole bioimpedance including muscles, fat, blood and other tissue.
With inner electrodes 3 and 4, the resulting voltage drop $V_\mathrm{M}$ over the bioimpedance is detected.
As described before, changes in the half-cell voltages caused by motion artifacts do not affect the bioimpedance measurement because of the frequency separation used.

In real applications, such as prosthetics, dry electrodes are preferably used.
Since these electrodes, in contrast to gel electrodes, do not contain electrolytes between the electrode and skin, they behave primarily capacitively and result in significantly higher values of $|Z_\mathrm{E}|$, typically in the $\text{k}\Omega$-range \cite{Dryelectrodesforbioimpedancemeasurementsdesigncharacterizationandcomparison, Bosnjak2017}.
It is conceivable that these conditions also affect technical implementations of the equivalent circuit in Fig. \ref{BioZ_ESB}.
Therefore, this section focuses on the influence of these electrode-skin impedances on real myographic impedance measurements.

\begin{figure}[!t]
	\centering
	\includegraphics[width=0.4\textwidth]{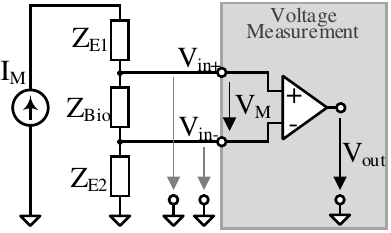}
\caption{Simplified equivalent circuit diagram of a bioimpedance measurement assuming infinitely high input impedances for the voltage measurement.}
	\label{EinflussESIsVCCM}
\end{figure}

The input impedances of the voltage measurement circuit can be chosen to be much higher than the values of the considered electrode-skin impedances, allowing us to neglect $Z_\mathrm{E3}$ and $Z_\mathrm{E4}$.
This results in a simplified equivalent circuit, as shown in Fig. \ref{EinflussESIsVCCM}.

In this circuit diagram, the electrode-skin impedances $Z_\mathrm{E1}$ and $Z_\mathrm{E2}$, as well as the bioimpedance $Z_\mathrm{Bio}$, are combinations of their passive elements, as shown above in Fig \ref{BioZ_ESB}.

The voltage measurement is represented by a real differential amplifier.
Since the amplifier is not ideal, it amplifies not only the wanted differential component of the input signal

\begin{align}
V_\mathrm{M}=V_\mathrm{in+}-V_\mathrm{in-}
\end{align}

but also the undesired common mode voltage

\begin{align}
V_\mathrm{CM}=\frac{V_\mathrm{in+}+V_\mathrm{in-}}{2}.
\end{align}

Assuming the differential amplification to be $A_\mathrm{D}=1$ and considering the common mode gain $A_\mathrm{CM}$, the output voltage of the amplifier circuit is

\begin{align}
V_\mathrm{out}&=A_\mathrm{D} V_\mathrm{M} + A_\mathrm{CM} V_\mathrm{CM}\\ \nonumber
&=V_\mathrm{in+}-V_\mathrm{in-}+ A_\mathrm{CM}\frac{V_\mathrm{in+}+V_\mathrm{in-}}{2}\\ \nonumber
&=I_\mathrm{M}\cdot\left( Z_\mathrm{Bio}+ A_\mathrm{CM}\frac{Z_\mathrm{Bio}+2\cdot Z_\mathrm{E2}  }{2}       \right).
\end{align}

\begin{figure}[!t]
	\centering
	\includegraphics[width=0.3\textwidth]{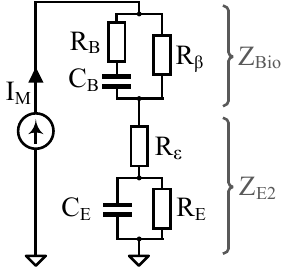}
\caption{Detailed equivalent circuit diagram of the bioimpedance and the negative current electrode-skin impedance.}
	\label{EIMesb}
\end{figure}

Accordingly, an incorrectly measured impedance value $Z_\mathrm{M}$ of

\begin{align}
\label{equ:ImpedaneIncorrect}
Z_\mathrm{M}&=Z_\mathrm{Bio}+A_\mathrm{CM}\frac{Z_\mathrm{Bio}+2\cdot Z_\mathrm{E2}}{2}\\ \nonumber
&=\left(1+\frac{A_\mathrm{CM}}{2} \right) Z_\mathrm{Bio}  +  A_\mathrm{CM} Z_\mathrm{E2}
\end{align}

is determined.
In (\ref{equ:ImpedaneIncorrect}), it can be seen that the impedance $Z_\mathrm{E1}$ has no influence on $Z_\mathrm{M}$ and can therefore be ignored.
For a more detailed analysis, the relevant components of $Z_\mathrm{Bio}$ and $Z_\mathrm{E2}$ in this context are shown in Fig. \ref{EIMesb}.

%

Applying these variables to (\ref{equ:ImpedaneIncorrect}) results in the complex representation

\begin{align}
\text{\begin{small}
	$
Z_\mathrm{M}\negthickspace=\negthickspace\left( \negthickspace 1\negthickspace+\negthickspace \frac{\negthickspace A_\mathrm{CM}}{2}  \negthickspace  \right) \negthickspace  \frac{R_{\beta} \negthickspace + \negthickspace \omega^2\negthickspace R_{\beta} \negthickspace R_\mathrm{B} \negthickspace C_\mathrm{B}^2 \negthickspace \left( \negthickspace R_{\beta} \negthickspace + \negthickspace R_\mathrm{B} \negthickspace \right) \negthickspace}{1 \negthickspace + \negthickspace \omega^2 C_\mathrm{B}^2\left(R_{\beta} \negthickspace + \negthickspace R_\mathrm{B}\right)^2}  \negthickspace   +  \negthickspace   A_\mathrm{CM} \negthickspace  \frac{R_{\epsilon} \negthickspace + \negthickspace  \textcolor{blue}{ R_\mathrm{E}} \negthickspace + \negthickspace \omega^2 \negthickspace R_{\epsilon} \negthickspace R_\mathrm{E}^2 \negthickspace C_\mathrm{E}^2}{1 \negthickspace + \negthickspace \omega^2 \negthickspace R_\mathrm{E}^2 \negthickspace C_\mathrm{E}^2}
	$
	\end{small}}\nonumber
\label{equ:ZmessEIMkomplexPart1}
\end{align}
\vspace{-3mm}
\begin{align}
\text{\begin{small}
	$
-\negthickspace \mathrm{j} \negthickspace \left( \negthickspace    \left( \negthickspace  1 \negthickspace + \negthickspace \frac{A_\mathrm{CM}}{2} \negthickspace \right) \negthickspace  \frac{\omega \negthickspace R_{\beta}^2 \negthickspace C_\mathrm{B}}{1\negthickspace +\negthickspace \omega^2 \negthickspace C_\mathrm{B}^2 \negthickspace \left( \negthickspace R_{\beta}\negthickspace + \negthickspace R_\mathrm{B} \negthickspace \right)\negthickspace ^2 }                                     \negthickspace + \negthickspace  A_\mathrm{CM} \negthickspace \frac{\omega \negthickspace  \textcolor{blue}{R_\mathrm{E}^2 \negthickspace C_\mathrm{E}}}{1 \negthickspace+ \negthickspace \omega^2 \negthickspace R_\mathrm{E}^2 \negthickspace C_\mathrm{E}^2}                              \right)
	$,
	\end{small}}
\end{align}

in which both the real and imaginary parts consist of two addends, caused by $Z_\mathrm{Bio}$ (left) and $Z_\mathrm{E2}$ (right).
The parallel circuit, consisting of $R_\mathrm{E}$ and $C_\mathrm{E}$ in Fig. \ref{EIMesb}, can be interpreted as a simplified nonideal capacitance, which again depends on the electrode contact area $\mathrm{A}$.
If mechanical disturbances lead to a reduction $k$ of this area according to

\begin{align}
A\Rightarrow\frac{A}{(1+k)},
\end{align}

$R_\mathrm{E}$ is increased by the factor $(1 + k)$.
$C_\mathrm{E}$ decreases according to the equation for a plate capacitor $C=\epsilon A/D$ with distance of D between the plates.
Although $R_\mathrm{\epsilon}$ is also influenced by movement, it is assumed to affect the contact impedance of dry electrodes much less than the parallel circuit $R_\mathrm{E}||C_\mathrm{E}$ and is therefore neglected.
Replacement of the variables in (\ref{equ:ZmessEIMkomplexPart1}) according to

\begin{align}
R_\mathrm{E} \Rightarrow R_\mathrm{E} \cdot (1+k)
\label{equ:ZmessEIMSubst}
\end{align}
\begin{align}
C_\mathrm{E} \Rightarrow  C_\mathrm{E}/(1+k)
\label{equ:ZmessEIMSubst2}
\end{align}

only affects the blue marked parts.
In other parts of the equation, the factors cancel each other out.
Thus, an electrode lift-off caused by motion artifacts increases the real part of $Z_\mathrm{M}$ according to

\begin{align}
\Delta Re\{Z_\mathrm{M}\} = A_\mathrm{CM} \cdot \frac{k R_\mathrm{E}}{1+\omega^2 R_\mathrm{E}^2C_\mathrm{E}^2}
\label{equ:DeltaReal}
\end{align}

and the imaginary part according to

\begin{align}
\Delta Im\{Z_\mathrm{M}\} = -A_\mathrm{CM} \cdot \frac{k \omega R_\mathrm{E}^2 C_\mathrm{E}}{1+\omega^2 R_\mathrm{E}^2C_\mathrm{E}^2}
\label{equ:DeltaImag}.
\end{align}

Therefore, an increase in the measured impedance magnitude over the whole frequency range is to be expected.
Inserting realistic values of dry electrodes from the literature into these equations ($C_\mathrm{E}\approx 150~\text{pF}...50~\text{nF}$; $R_\mathrm{E}\approx 100~\text{k}\Omega...1~\text{M}\Omega$ \cite{Heikenfeld2018, Dryelectrodesforbioimpedancemeasurementsdesigncharacterizationandcomparison}), it can be seen that $|\Delta Re\{Z_\mathrm{M}\}|\ll |\Delta Im\{Z_\mathrm{M}\}|$.
As described in the literature and based on previous measurements, the phase shift of $Z_\mathrm{M}$ is typically in the range of $\phi(Z_\mathrm{M})\approx0^{\circ}...-30^{\circ}$, caused by the dominant influence of $Z_\mathrm{Bio}$ \cite{Rutkove2009,CombiningBioimpedanceandEMGMeasurementsforReliableMuscleContractionDetection}.
Thus, the impedance changes mentioned above, which primarily affect the imaginary part, always lead to more negative phase shifts of the measured impedance $Z_\mathrm{M}$.
Although the numerical values of the explained relationships depend on $\omega$, they do not change their characteristic direction as a function of the frequency.
In contrast, the PhaseX Effect described before shows completely different behavior over frequency, which cannot be the result of motion artifacts.

\begin{figure}[!t]
	\centering
	\includegraphics[width=0.48\textwidth]{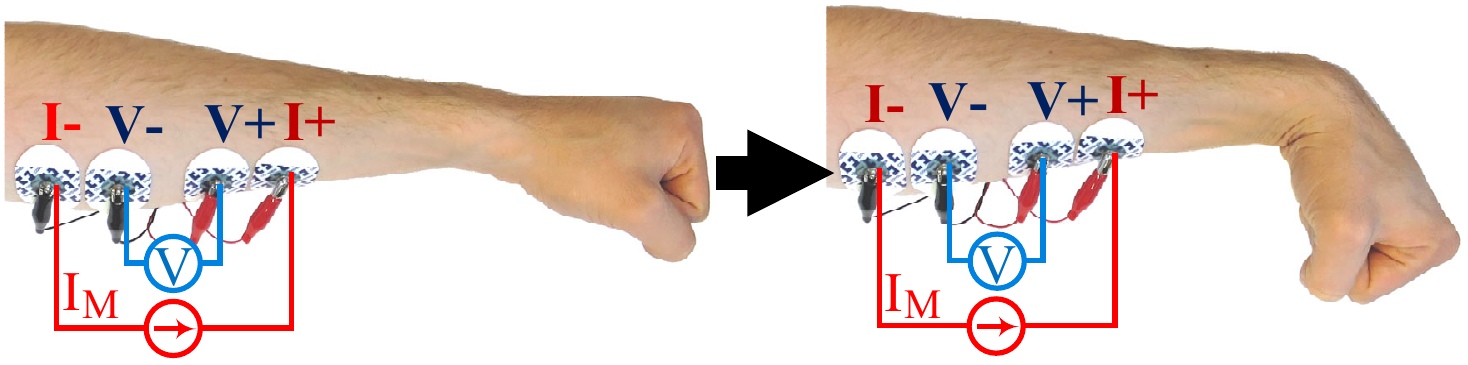}
\caption{Electrode placement for measuring muscle contractions during wrist flexion.}
	\label{UnterarmKontraktion}
\end{figure}

\subsection{Measurement Setup}
For further investigation of the proposed effect, first subject measurements were performed.
A bioimpedance spectrometer was used that covers the typical measurement frequency range and ensures electrical safety \cite{Ahighaccuracybroadbandmeasurementsystemfortimeresolvedcomplexbioimpedancemeasurements}.
Muscle contractions on the forearm were measured by means of a 4-terminal setup using four Ag/AgCl hydrogel electrodes (H92SG from Kendall).
The electrodes were evenly spaced approximately $30~\text{mm}$ apart, and they were placed above the muscle flexor pollicis longus, as shown in Fig. \ref{UnterarmKontraktion}.
The excitation current was $I_\mathrm{M} = \text{1.25~mA}$, and the signal form was a linear chirp in the frequency range of $\text{24.4...390~kHz}$.
Due to the system's ability to acquire up to 3480 complex impedance spectra per second, each measurement took less than $300~\mu s$ to obtain.
Since it can be assumed that the actual impedance changes due to physiological effects take much longer, the acquisition duration of the impedance spectra can be neglected.


\section{Results and Discussion}
In this section, the measurement results from three subjects are presented and discussed.
Afterward, a setup that was used to synchronize offline measurements with a simultaneously captured video of the performed arm motions is proposed.
The corresponding video file can be downloaded from \href{https://ieeexplore.ieee.org}{https://ieeexplore.ieee.org}.

\subsection{Subject Measurements} \label{sec:SeubjectMeasurements}
\begin{figure*}[!t]
	\centering
	\includegraphics[width=1\textwidth]{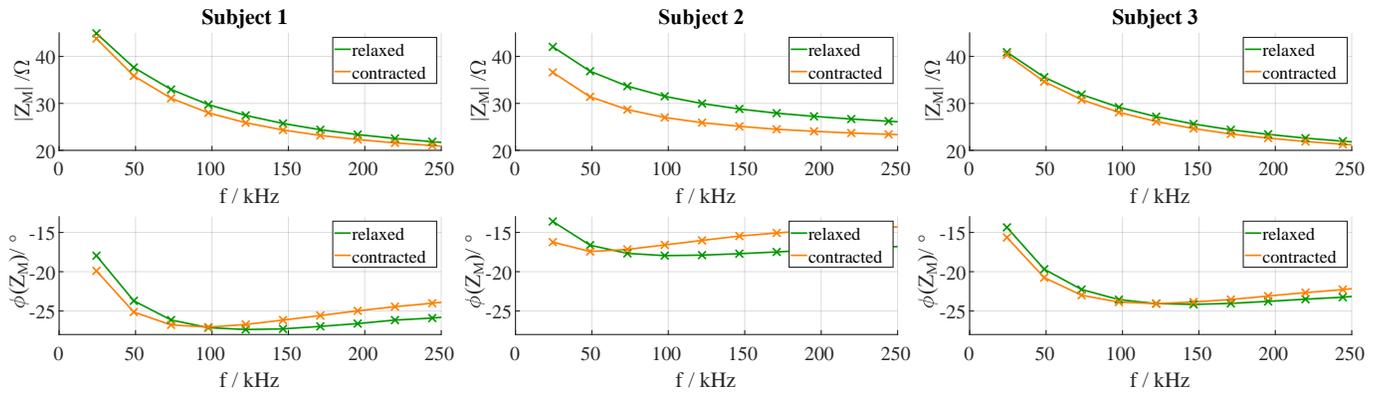}%
\caption{Measured complex bioimpedance spectra from three subjects during relaxation (green) and contraction (orange) of the corresponding forearm muscle.}
	\label{SubjectStudy}
\end{figure*}

Using the previously described measurement setup, the bioimpedances at the forearm were determined for three young subjects in a relaxed muscle state.
Subsequently, the subjects contracted their muscles as described and shown in Fig. \ref{UnterarmKontraktion} so that the wrist was bent.
The acquired complex frequency responses from both conditions for the three subjects' forearms are shown in Fig. \ref{SubjectStudy}. As explained above, $Z_\mathrm{M}$ represents the actual bioimpedance $Z_\mathrm{Bio}$, including the technical influences of the measurement device (see (\ref{equ:ImpedaneIncorrect})).
Since the results are most interesting in the frequency range below $\text{250~kHz}$, the data for higher frequencies are not plotted.

In the upper plots of all three subjects, it can be seen that the magnitude responses have behaviors very similar to those of the model, as shown in Fig. \ref{Modell}. The impedances corresponding to the relaxed state (green) begin at approximately $40~\Omega$ and decrease with frequency due to the capacitive behavior of the bioimpedances. When the muscle is contracted, the whole magnitude response for all subjects is shifted to lower impedance values, as shown in the orange plots. This effect is similar to the presented model also shown in Fig. \ref{Modell} and does not significantly depend on the frequency of the excitation signal.

In the lower plots, the corresponding phase responses are presented. In the green plots, acquired when the muscle was relaxed, the typical electrical characteristic of the bioimpedance equivalent circuit, shown in Fig. \ref{MuscleFibreESB}, can be recognized.
Since the equivalent circuit has a resistive behavior for $f\rightarrow0$ and $f\rightarrow \infty$, the minimum of the phase response is in between.
For bioimpedances, this minimum is typically in the range of tens to low hundreds of kilohertz \cite{BioelectricityandBioimpedanceBasics}.
These minima can also be seen in all three green plotted phase responses.
The orange phase plots correspond to the contracted muscle states and also have this characteristic minimum. Compared to those of the green plots, the specific frequencies of the minima are lower, whereas the shapes of the phase responses are not significantly changed, especially in the plots of subjects 1 and 3.
This behavior corresponds to the theoretical idea presented in section \ref{sec:EIMphaseX} of this work and evokes the PhaseX Effect in all subject measurements.
The crossings are located at $f_\mathrm{PhaseX,1}\approx 100~\text{kHz}$, $f_\mathrm{PhaseX,2}\approx 65~\text{kHz}$ and $f_\mathrm{PhaseX,3}\approx 125~\text{kHz}$.
As described earlier, this effect is promising as a reliable marker for detecting actual muscle contractions and for differentiating them from external disturbances.

Even if these first subject measurements show the expected behavior and therefore support the idea of this work, the data must be interpreted carefully.
The actual electrical characteristics of the bioimpedances are much more complex than those of the assumed equivalent circuit. The complexity includes not only the electrical behavior of the tissue but also that of the geometrical tissue distributions and the corresponding neglected anisotropy. In addition, it must also be considered that the skin, fat and body fluids affect the measured bioimpedance values. The influence of other geometrical changes of these tissues due to muscle contractions on the results shown cannot be excluded decisively.
Therefore, further subject studies are necessary to ensure the actual suitability of this effect. For their intended use in prosthesis control, these measurements should be obtained from different muscle regions in combination with common EMG signals \cite{CombiningBioimpedanceandEMGMeasurementsforReliableMuscleContractionDetection,Applicationofsinglewirelessholter}. These measurements should also include experienced users of myoelectrical prostheses.

\subsection{Synchronized Video}
\begin{figure}[!t]
	\centering
	\includegraphics[width=0.48\textwidth]{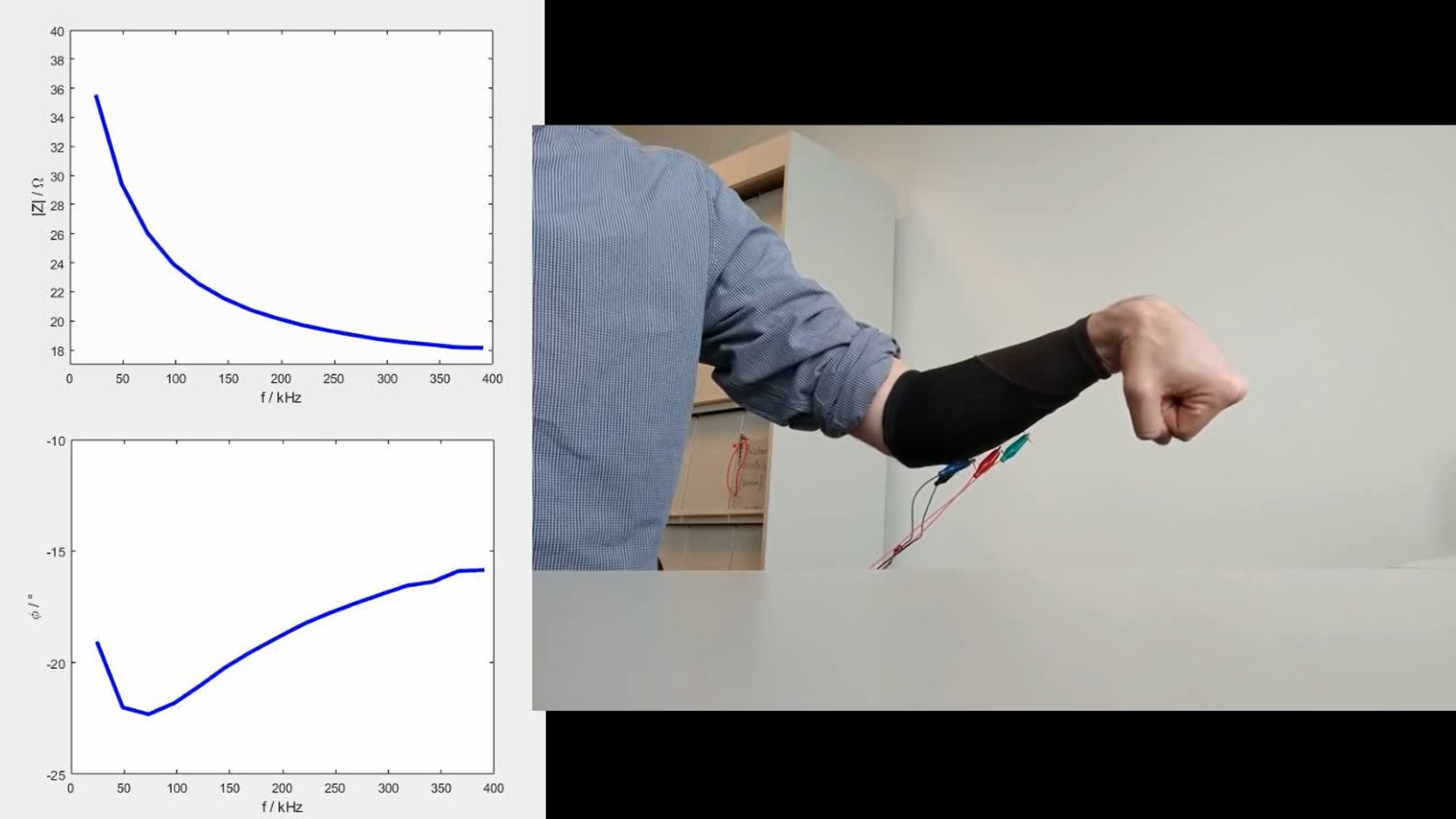}
\caption{Screenshot of the acquired video, which combines the measurement data and the actual performed muscle contractions. The video is available at \href{https://ieeexplore.ieee.org}{https://ieeexplore.ieee.org}.}
	\label{VideoScreenshot}
\end{figure}

For better comparison of the signal characteristics between muscle contractions and motion artifacts, the previous measurements were repeated during additional video capturing.
To generate more realistic conditions found in prosthesis applications, dry gold electrodes with diameters of $15~\text{mm}$ were used \cite{Dryelectrodesforbioimpedancemeasurementsdesigncharacterizationandcomparison}.
These electrodes were placed via a forearm sleeve at the same position as the electrodes used in the subject study described before.
The setup can be seen in the photograph on the right side of Fig. \ref{VideoScreenshot}.
The bioimpedance measurement was performed with a sampling rate of 3480 complex impedance spectra per second, in contrast to the video, which has a time resolution of $30~\text{fps}$.
To smooth the data, a moving average filter with a length of 500 samples, corresponding to approximately $144~\text{ms}$, was applied to the bioimpedance spectra.
Afterward, the signals were plotted and manually synchronized with the video from the camera. Due to this kind of synchronization, time shifts of up to $500~\text{ms}$ between the presented data and the shown muscle contraction video could occur.
However, for qualitative analysis of the PhaseX Effect, the observed synchronicity was sufficient. The subject was asked to perform several motions with the wrist and to generate motion artifacts by shaking the arm.

The combined video file is available at \href{https://ieeexplore.ieee.org}{https://ieeexplore.ieee.org}. The effects described earlier can clearly be seen in the magnitude and phase plots. In particular, the expected shift of the minimum in the phase response is very promising for real-time applications. In this experiment, the measurement results are very robust to the generated motion artifacts, even when dry electrodes are used.

\subsection{Statistical Analysis}
\begin{figure}[!t]
	\centering
	\includegraphics[width=0.32\textwidth]{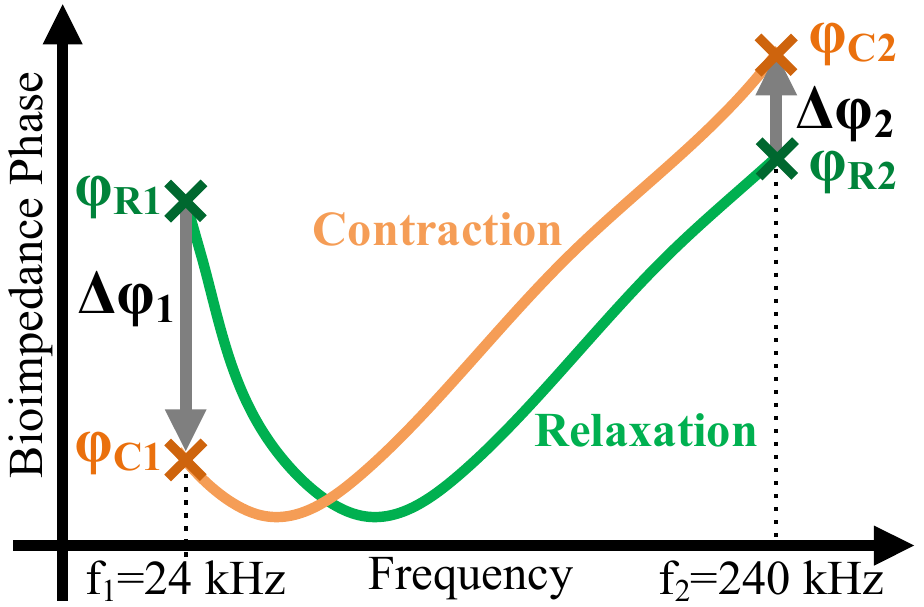}
\caption{Example principle for detecting a muscle contraction via the phase responses of the bioimpedance.}
	\label{StudyPrinciple}
\end{figure}

The measurement results shown in Fig. \ref{SubjectStudy} represent only one single muscle contraction per subject. To provide additional meaningful information, statistical analysis was performed. One subject was asked to repeat the wrist flexion (see Fig. \ref{UnterarmKontraktion}) 100 times under the same measurement conditions as described in section \ref{sec:SeubjectMeasurements}. The durations of the contractions and relaxations were 2 s each. In a second setup, the subject contracted the muscles of the upper forearm 100 times to perform wrist extensions. The electrodes used to measure the bioimpedance were placed on the upper forearm in this measurement configuration. For the simple detection of muscle contractions, the change in the phase response was analyzed. As shown in Fig. \ref{StudyPrinciple}, only two frequency points ($f_1=24~\mathrm{kHz}$, $f_2=240~\mathrm{kHz}$) from the phase responses during relaxation ($\varphi_\mathrm{R1}$, $\varphi_\mathrm{R2}$) and contraction ($\varphi_\mathrm{C1}$, $\varphi_\mathrm{C2}$) were used.

According to the introduced PhaseX Effect, a muscle contraction fulfills the condition presented in (\ref{equ:StatisticStudy}).

\begin{align}
\frac{ \varphi_\mathrm{C1}-\varphi_\mathrm{R1}}{\varphi_\mathrm{C2}-\varphi_\mathrm{R2}} = \frac{\Delta \varphi_\mathrm{1}}{\Delta \varphi_\mathrm{2}} < 0
\label{equ:StatisticStudy}
\end{align}

\begin{figure}[!t]
	\centering
	\includegraphics[width=0.4\textwidth]{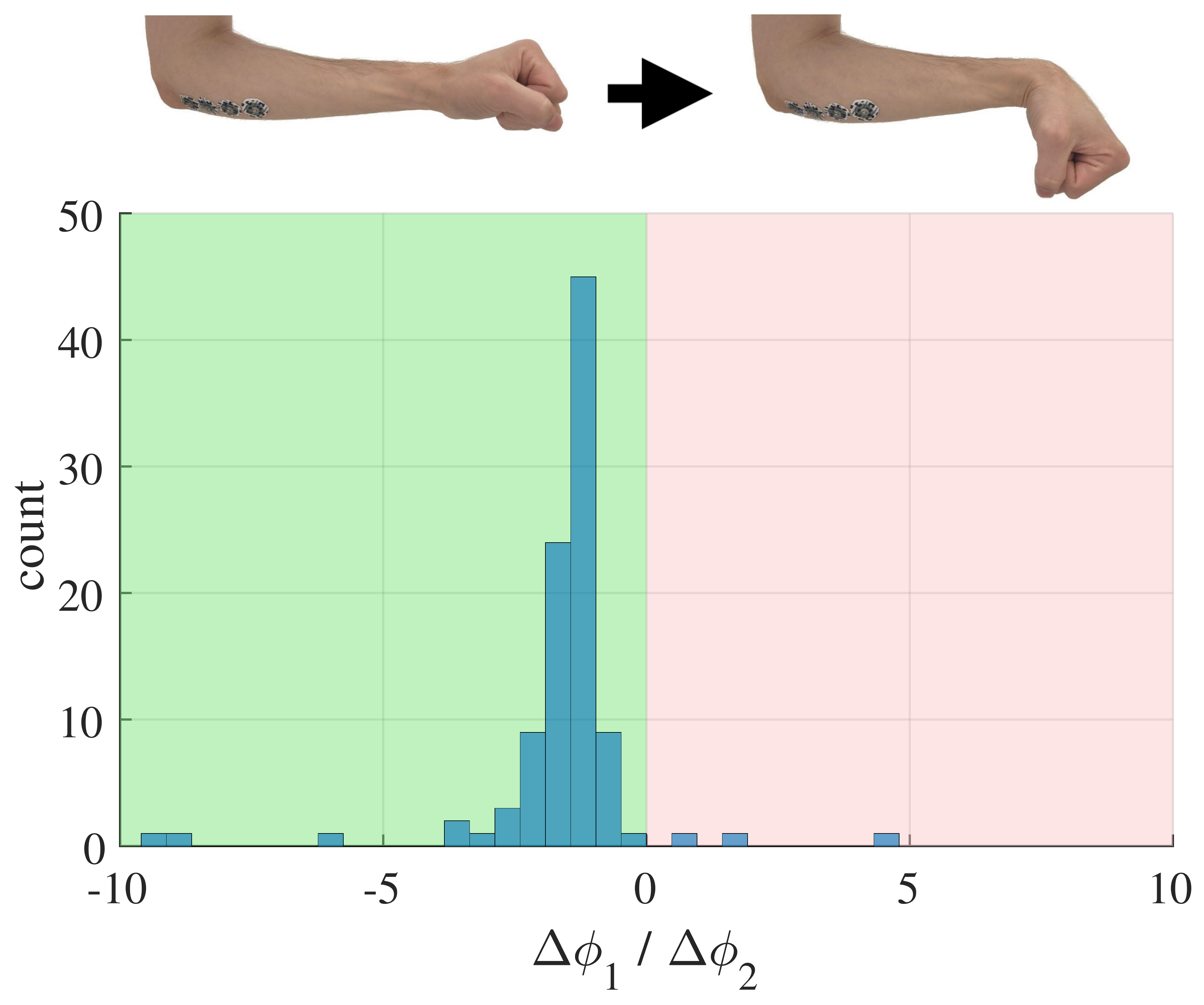}
\caption{Histogram representation of the detected muscle contractions during wrist flexion.}
	\label{StudyLowerSideResults}
\end{figure}

The corresponding detected values of the 100 contractions performed for the wrist flexions are shown as a histogram in Fig. \ref{StudyLowerSideResults}. It can be seen that 97 of the 100 contractions were recognized correctly with this simple signal processing approach. The mean value of this parameter was measured to be ${mean_\mathrm{flexion}\left(\frac{\Delta \varphi_\mathrm{1}}{\Delta \varphi_\mathrm{2}}\right)=~-1.6}$.

\begin{figure}[!t]
	\centering
	\includegraphics[width=0.4\textwidth]{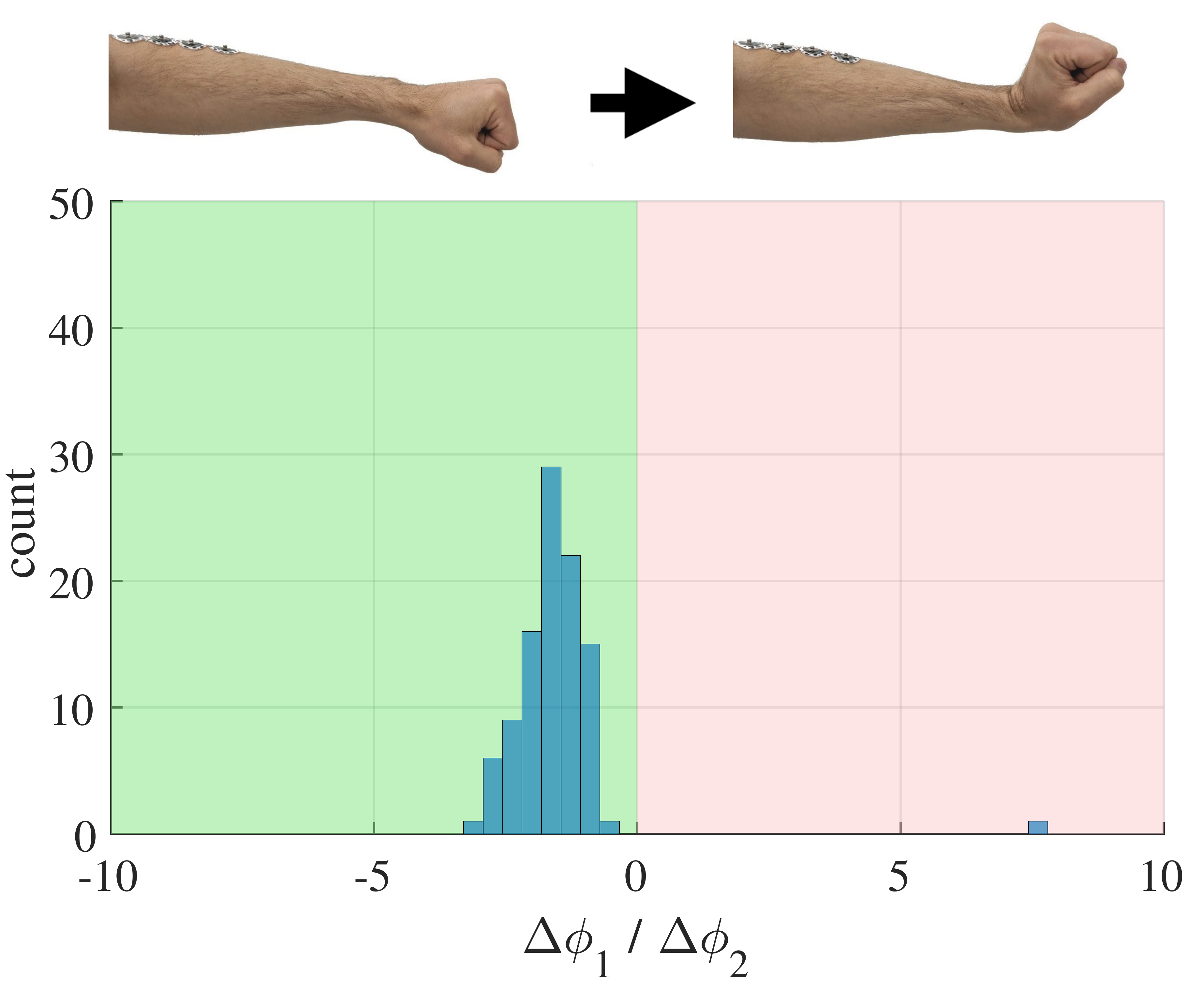}
\caption{Histogram representation of the detected muscle contractions during wrist extension.}
	\label{StudyUpperSideResults}
\end{figure}

In Fig. \ref{StudyUpperSideResults}, the corresponding histogram of the measurements performed during the wrist extensions is shown. For these measurements, only 1 of 100 muscle contractions was not detected correctly. The mean value was measured to be ${mean_\mathrm{extension}\left(\frac{\Delta \varphi_\mathrm{1}}{\Delta \varphi_\mathrm{2}}\right)=~-1.5}$.

Since neither the actual strength of the muscle contractions nor the angles of the wrist motions were captured, the quantitative results must be interpreted carefully. However, the measurements indicate the reliability of the approach.


\section{Conclusion}
In this work, we presented a novel approach for the reliable detection of muscle contractions via multi-frequency bioimpedance measurements.
In particular, the focus of the work was on the behavior of the phase response.
After explaining the idea of the phase sensitivity of certain muscle contractions, a theoretical analysis of the effect of typical electrode disturbances was presented.
It was shown that changes in the electrode contacts due to motion artifacts cannot produce the same signal characteristics as an actual muscle contraction.
Therefore, the approach of measuring the phase response is promising for reliable signal analysis.

Subject measurements on the forearm showed very similar impedance behaviors as the results of the theoretical model.
The particular shift of the phase response when the corresponding muscle is contracted can be used as a new marker that can be beneficial in several applications, such as prosthesis control.
Since the phase plots obtained during both relaxation and contraction always cross, elementary signal processing algorithms are sufficient for generating high reliable results.
One example approach was presented in combination with statistical analysis.
The combination of a reliable marker and simple signal analysis not only would be of interest for biomedical applications but can also be useful in human-computer interactions.
Based on the findings of this work, significantly simplified measurement systems can be developed in the future that enable more extensive subject studies.

\section*{Acknowledgment}
The authors would like to thank Steffen Kaufmann for providing the bioimpedance spectrometer and Sebastian Hauschild for performing the preliminary investigations.


\bibliographystyle{IEEEtran}
\bibliography{bib}

\end{document}